\documentclass{article}

\usepackage{PRIMEarxiv}

\usepackage[utf8]{inputenc} 
\usepackage[T1]{fontenc}    
\usepackage{hyperref}       
\usepackage{url}            
\usepackage{booktabs}       
\usepackage{amsfonts}       
\usepackage{nicefrac}       
\usepackage{microtype}      
\usepackage{lipsum}
\usepackage{fancyhdr}       
\usepackage{graphicx}       
\graphicspath{{media/}}     

\title{Ring Current Proton Decay Timescales Derived from Van Allen Probe Observations
\thanks{\textit{\underline{Citation}}: 
\textbf{Wang, S. and Li, J. (2023), Ring Current Proton Decay Timescales Derived from Van Allen Probe Observations, https://doi.org/10.48550/arXiv.2307.08907. This study was presented at the Geospace Environment Modeling (GEM) 2023 meeting in San Diego.}} 
}

\author{
  Stephanie Wang \\
  Stanford University Online High School\\
  \texttt{stephyxwang@gmail.com} \\
   \And
  Jinxing Li \\
  University of California, Los Angeles \\
}

\begin{document}
\maketitle

\begin{abstract}
The Earth’s ring current is highly dynamic and is strongly influenced by the solar wind. The ring current alters the planet’s magnetic field, defining geomagnetic storms. In this study, we investigate the decay timescales of ring current protons using observations from the Van Allen Probes. Since proton fluxes typically exhibit exponential decay after big storms, the decay time scales are calculated by performing linear regression on the logarithm of the fluxes. We found that in the central region of the ring current, proton decay timescales generally increase with increasing energies and increasing L-shells. The ~10s keV proton decay timescales are about a few days, while the ~100 keV proton decay time scale is about ~10 days, and protons of 269 keV have decay timescales up to ~118 days. These findings provide valuable insights into the ring current dynamics and can contribute to the development of more accurate ring current models.

\end{abstract}


\section{Introduction}
The Earth's magnetized space, known as the magnetosphere, is the region around our planet where its magnetic field dominates that of the solar wind. Observations from various satellites reveal intimate connections between Earth's magnetosphere and the solar wind, which is the direct driver of geomagnetic storms and the variation of geomagnetic field. Emitted by the sun, solar wind is a stream of charged particles (mostly protons and electrons). During periods of enhanced solar wind activity, the solar wind exerts pressure on Earth's magnetosphere, intensifying the magnetic field within the magnetosphere, leading to energetic particle injection from the magnetotail and the formation of the ring current.

The Earth’s ring current is an electric current encircling the Earth at the height of ~10,000 km to ~40,000 km near the magnetic equatorial plane. It consists of energetic ions, primarily protons, trapped by the geomagnetic field. The ring current dynamics causes global magnetic field variation and distortion, known as geomagnetic storms. Geomagnetic storms and ring current dynamics are driven by solar activities, either a solar coronal mass ejection (CME) or a co-rotating interaction region (CIR) which is a high-speed solar wind originating from a coronal hole \cite{Borovsky2006}. Geomagnetic storms lead to a variety of associated effects in geospace including radiation belt enhancement \cite{Kivelson1995}. After the geomagnetic storm main phase, ring current proton fluxes exhibit exponential decay, and the geomagnetic field is subsequently recovered.

In this study, we calculate the ring current proton decay timescales at representative energies and different L-shells (the geocentric distances projected to the magnetic equator in units of Earth Radii) using Van Allen Probe observations. The present study aims to gain a deeper understanding of the dynamics of ring current protons, providing insights for future space missions and facilitating a more profound understanding of the universe. The energy-dependent decay timescales at different altitudes are useful for developing more accurate ring current models.

\section{Data Description}
The Van Allen Probes \cite{Mauk2013}, also known as the Radiation Belt Storm Probes (RBSP), were a pair of NASA spacecraft designed to study the inner magnetospheric environments, especially the radiation belt surrounding the Earth. Launched on August 30, 2012, the mission aimed to improve our understanding of the radiation belts, the two donut-shaped regions of highly energetic charged particles trapped by Earth's magnetic field, known for their potentially hazardous effects inflicted on satellites, spacecraft, and astronauts. The Van Allen Probe’s main objectives include exploring the dynamics of high-energy ring current particles including electrons and ions, and studying the associated space weather effects in order to provide crucial information for future space exploration endeavors and satellite operations.

This study mainly investigates proton fluxes measured by the RBSPICE (Radiation Belt Storm Probes Ion Composition Experiment) instrument \cite{Mitchell2013} onboard Van Allen Probes. The RBSPICE instruments measure the fluxes for different types of ions in the inner magnetosphere over a wide range of energies. Specifically, they measure proton fluxes from 7 keV to 598 keV, covering the main energies of the ring current \cite{Lanzerotti2016}. Note that protons of ~100 keV is the dominant contributor to plasma pressure especially during quiet time\cite{Yue2018}.

\section{Data Overview}

\begin{figure}
  \centering
  \includegraphics[width=12cm]{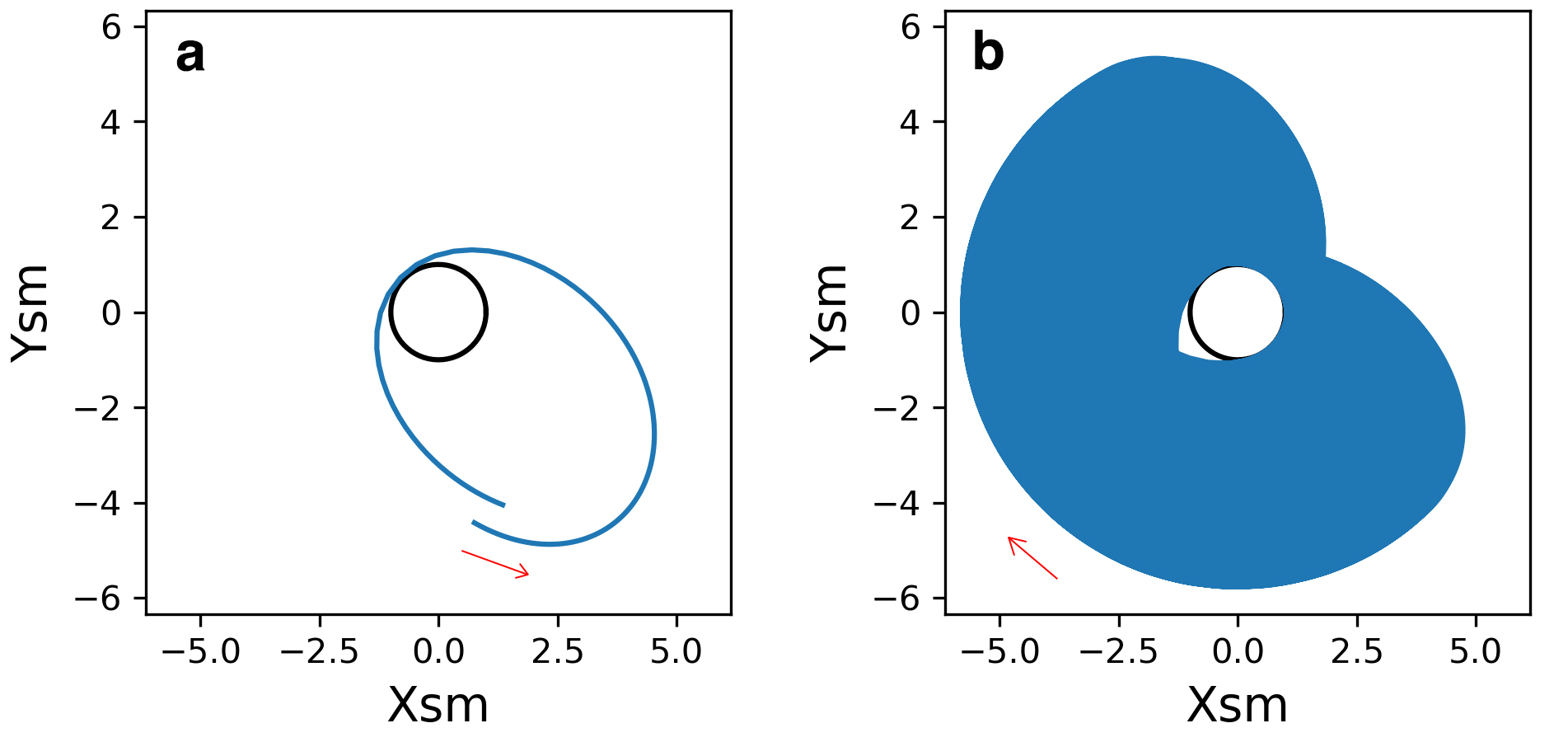}
  \caption{(a) The Van Allen Probe A path over an orbit period from 00 UT to 09 UT on 1 January 2018. The red arrow indicates the orbit direction. (b) The coverage of Van Allen Probe A orbit over the entire year of 2018. The red arrow indicates the orbit sweeping over local time.}
  \label{fig:fig1}
\end{figure}

Figure 1a displays the Van Allen Probe-A trajectory over an orbit period from 00 UT to 09 UT on 1 January 2018, showing an elliptical orbit with a period of about 9 hours. Figure 1b shows the orbital coverage over the entire year of 2018, showing that the satellite was sweeping westward over local time.  

An overview of the ring current proton fluxes during 2018 is presented in Figure 2. The proton fluxes are shown as a function of L shell at energies of 13 keV, 27 keV, 55 keV, 99 keV, 148 keV and 269 keV. Figure 2a shows the Sym-H index, which is the symmetric component of magnetic field disturbances at low latitudes on Earth’s surface, similar to the disturbance storm time (Dst) index. 

Figure 2 shows that ring current protons at low energies (~10s keV) are mainly observed at altitudes above L=4. Enhancement of these low energy protons is seen in response to each geomagnetic storm, including minor ones. On the other hand, protons with higher energies (~100s keV) generally peak at lower L shells and mainly respond to large storms. For example, the 269 keV protons around L=4 were only enhanced by the largest geomagnetic storm of the year in September, as seen in the Sym-H index. Moreover, Figure 2 demonstrates that ring current ions typically exhibit exponential decay after reaching peak values during geomagnetic storms.

\begin{figure}
  \centering
  \includegraphics[width=16cm]{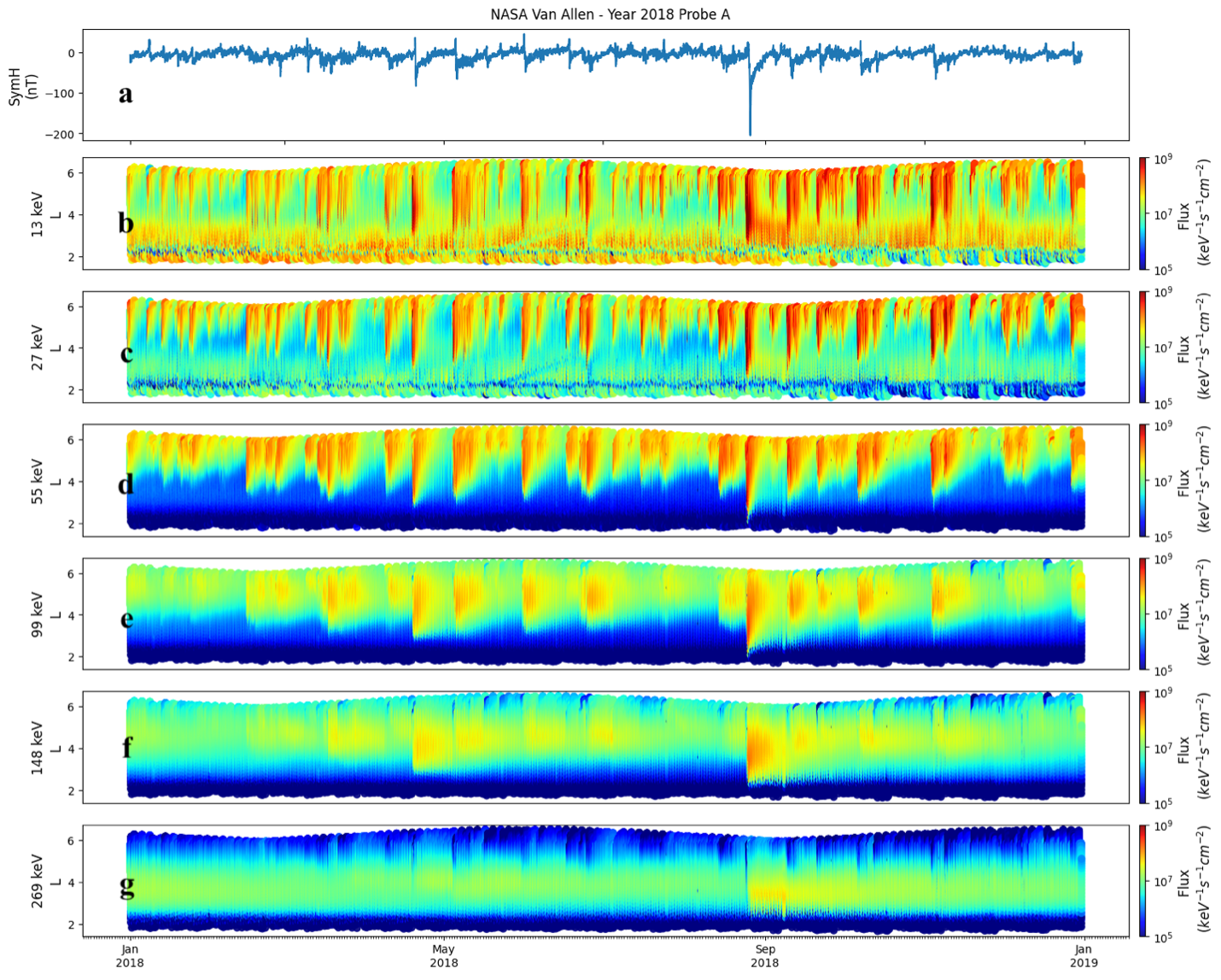}
  \caption{Geomagnetic indices and ion fluxes measured over the year of 2018. (a) The Sym-H index; (b-g) The proton fluxes as a function of L shell at energies of 13 keV, 27 keV, 55 keV, 99 keV, 148 keV, and 269 keV, respectively. Note that the proton fluxes of 13 keV and 27 keV are contaminated below L=3.}
  \label{fig:fig2}
\end{figure}

\section{Methodology}
Ring current ion fluxes typically exhibit exponential decay after reaching peak values due to injections and acceleration during geomagnetic storms. Therefore, the logarithm of flux can be fitted by a linear function of time. This study calculates the proton flux decay timescales using linear regression. Suppose we have a time sequence of ion flux $z_i$ measured at time $t_i$ during the decay phase measured by the satellite at a specific energy and a specific L shell, and $y_i$ = $log z_i$, then the slop of the fitting line can be calculated as
\begin{equation}
k = \frac{n\sum{y_{i}t_{i}} - \sum{y_i}\sum{t_i}}{n\sum{t_{i}^{2}} - (\sum{t_i})^2}
\end{equation}
The decay time scale is therefore  

\begin{equation}
\tau = -\frac{1}{k} = \frac{n\sum{t_{i}^{2}} - (\sum{t_i})^2}{\sum{y_i}\sum{t_i} - n\sum{y_{i}t_{i}}}
\end{equation}

\section{Case Studies}
To systematically investigate the ring current proton decay timescales, we show case studies at five selected energies, 27 keV, 55 keV, 99 keV, 148 keV and 269 keV, covering the main energies of the ring current. We illustrate cases that show clear exponential decay pattens, and calculate the energy-dependent proton decay timescales. 

\begin{figure}
  \centering
  \includegraphics[width=14cm]{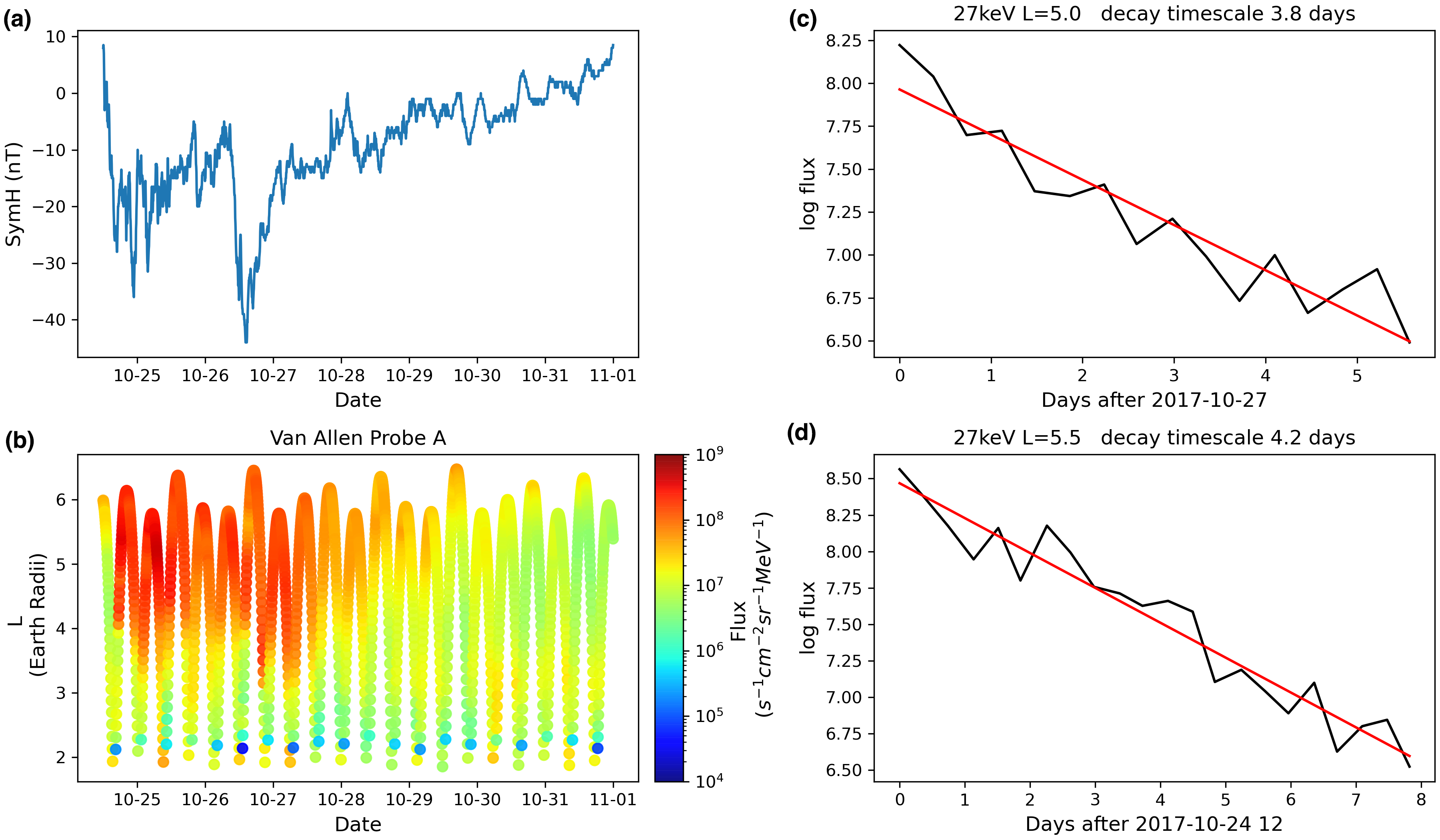}
  \caption{(Left) The geomagnetic Sym-H index and the 27 keV proton fluxes as a function of L shell measured from 24 October 2017 to 1 November 2017. (Right) Proton flux variations at L=5 and L=5.5, respectively, overplotted with the linear fitting lines. The decay timescales are derived from the slot of the linear regression.}
  \label{fig:fig3}
\end{figure}

Figure 3 shows our analysis for 27 keV protons. Figure 3a shows the Sym-H index from 24 October 2017 to 1 November 2017, indicating a geomagnetic storm recovery phase. Figure 3b shows the 27 keV proton flux versus L-shell measured by Van Allen Probe A along its orbit, indicating a rapid decay above L=4. Figure 3c shows the proton flux measured at L=5 in black line. The red line represents the linear fitting of the proton flux, and the decay timescale calculated using linear regression is 3.8 days for this case. Figure 3d shows the measured proton flux at L=5.5 and the linear fitting, indicating a decay timescale of 4.1 days, slightly longer than that at L=5.

We note that a few factors cause the data fluctuation along the linear fitting lines which represent exact exponential decay. 1) The satellite was at different latitudes each time when it traveled across a specific L shell, due to the unalignment between the Earth’s magnetic axis and its geographical axis. 2) Minor activities, including substorms, may have happened during the recovery phase of geomagnetic storms. Nevertheless, we see that the linear fitting can well represent the proton decay activity.

\begin{figure}
  \centering
  \includegraphics[width=14cm]{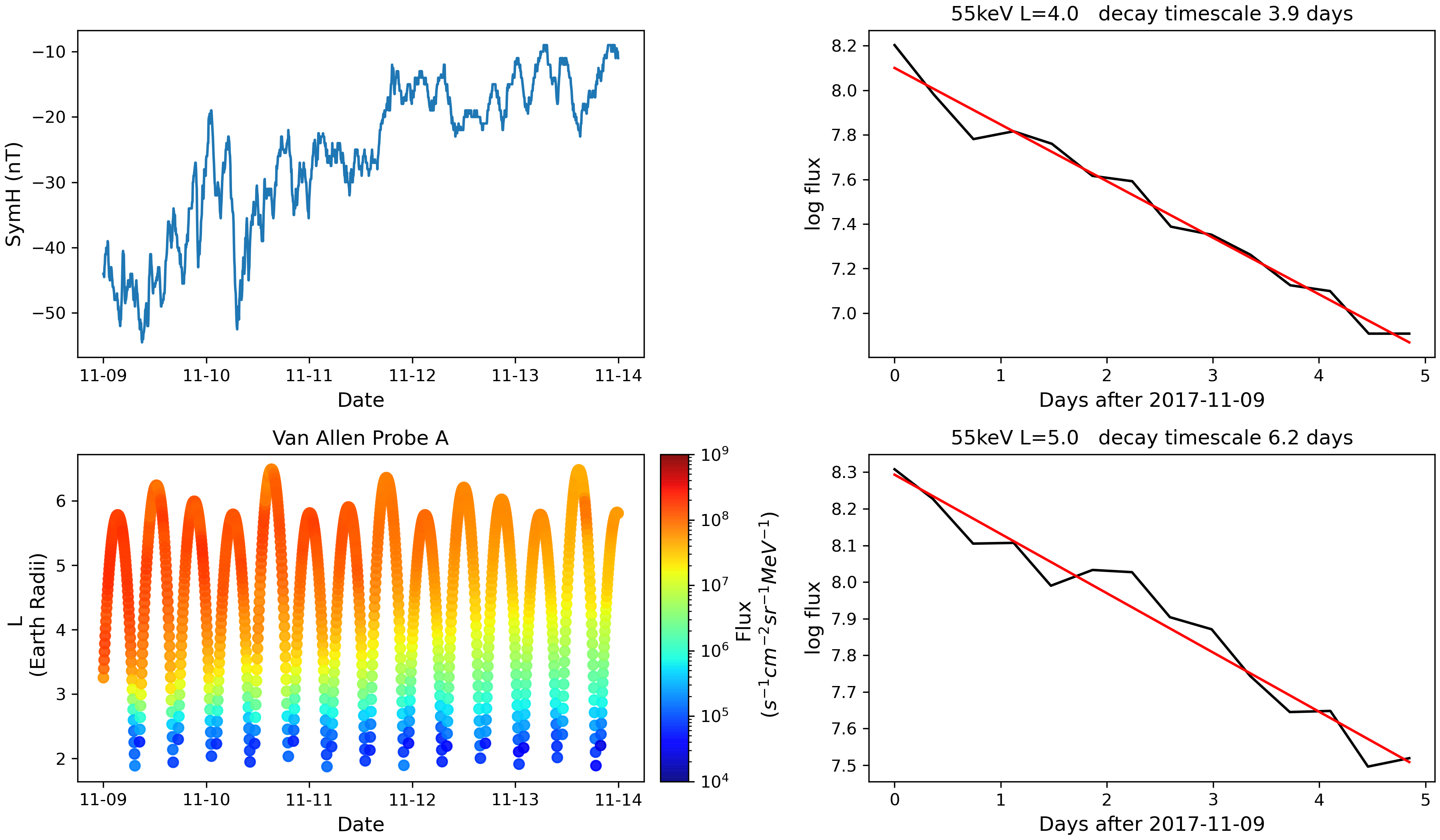}
  \caption{The same as Figure 3 but for 55 keV proton fluxes measured from 9 November 2017 to 14 November 2017. The proton decay timescales are calculated at L=4 and L=5, respectively.}
  \label{fig:fig4}
\end{figure}

Figure 4 shows the 55 keV proton flux observed from 9 November to 14 November in 2017. The linear regression indicates that the decay timescale is 3.9 days at L=4, and is 6.2 days at L=5 in this event.

\begin{figure}
  \centering
  \includegraphics[width=14cm]{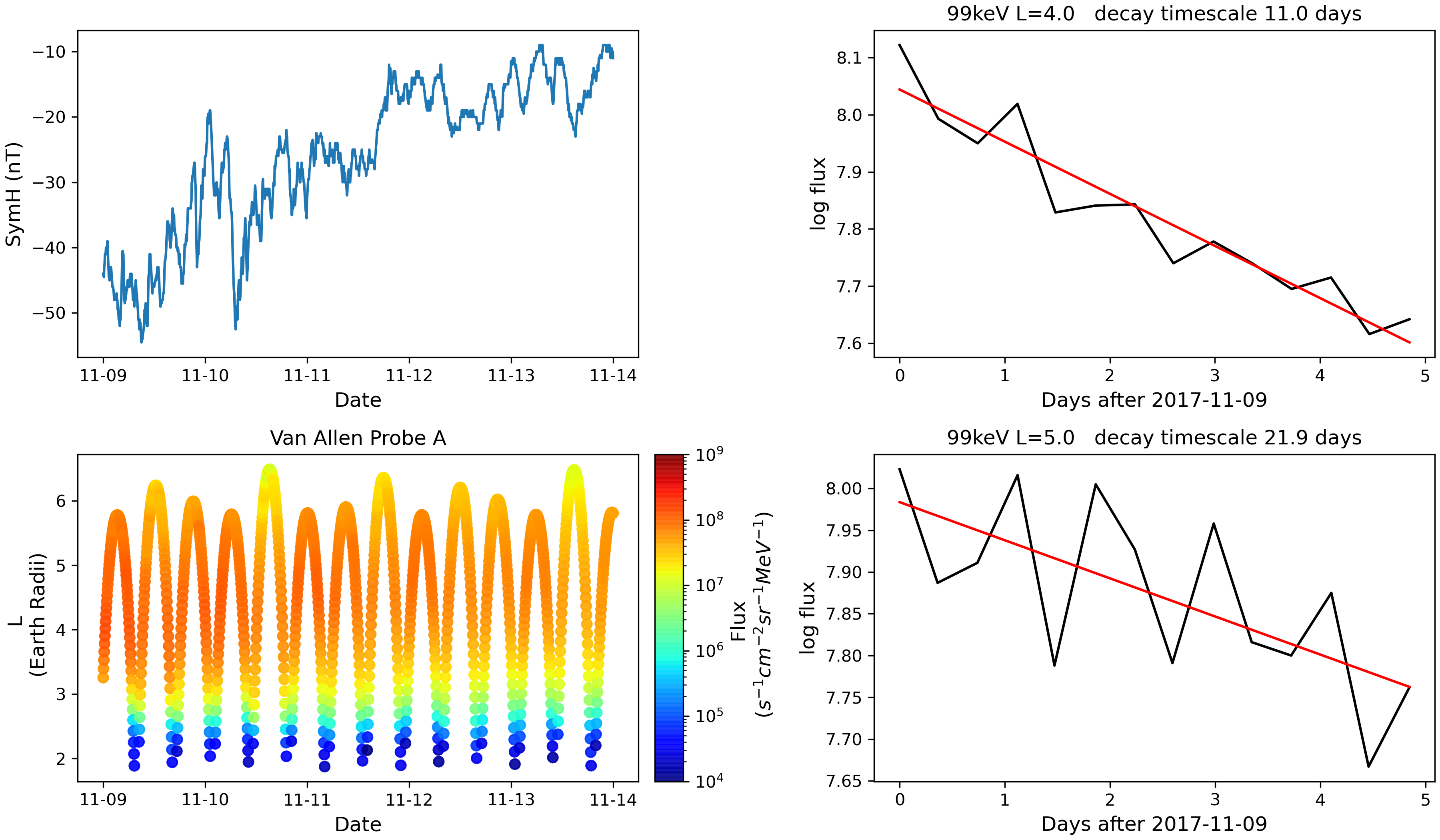}
  \caption{The same as Figure 4 but for 99 keV protons.}
  \label{fig:fig5}
\end{figure}

Figure 5 shows the satellite observation and linear regression from 9 November 2017 to 14 November 2017 for 99 keV protons. The proton decay timescale is 11.0 days at L=4, and is 21.9 days at L=5. 

\begin{figure}
  \centering
  \includegraphics[width=14cm]{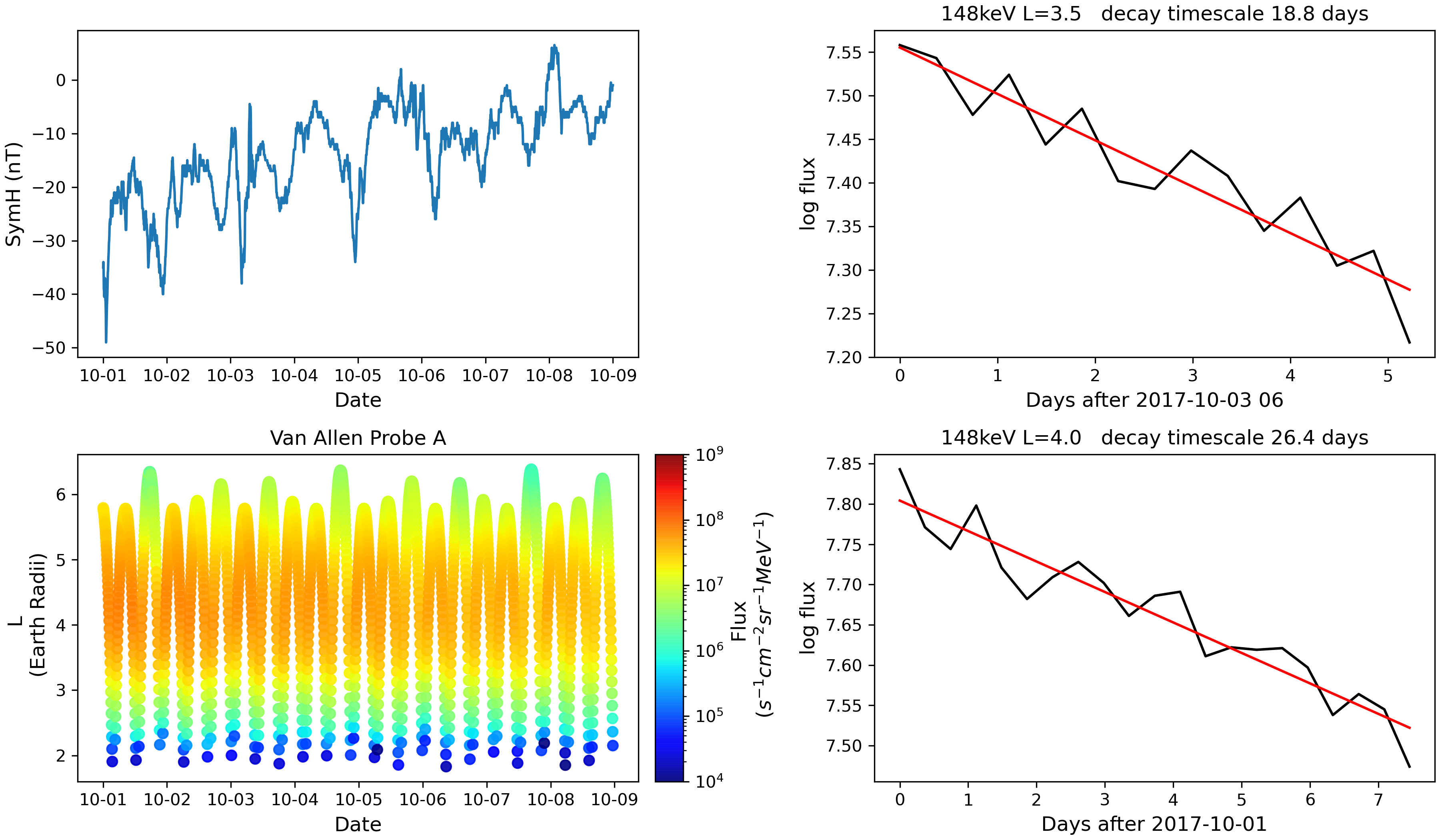}
  \caption{The same as Figure 3 but for 148 keV proton fluxes measured from 1 October 2017 to 9 October 2017, and the proton decay timescales are calculated at L=3.5 and L=4.}
  \label{fig:fig6}
\end{figure}

Figure 6 exhibits the study for 148 keV protons, and our calculation produces a decay timescale of 18.8 days at L=3.5, and a timescale of 26.4 days at L=4.

\begin{figure}
  \centering
  \includegraphics[width=14cm]{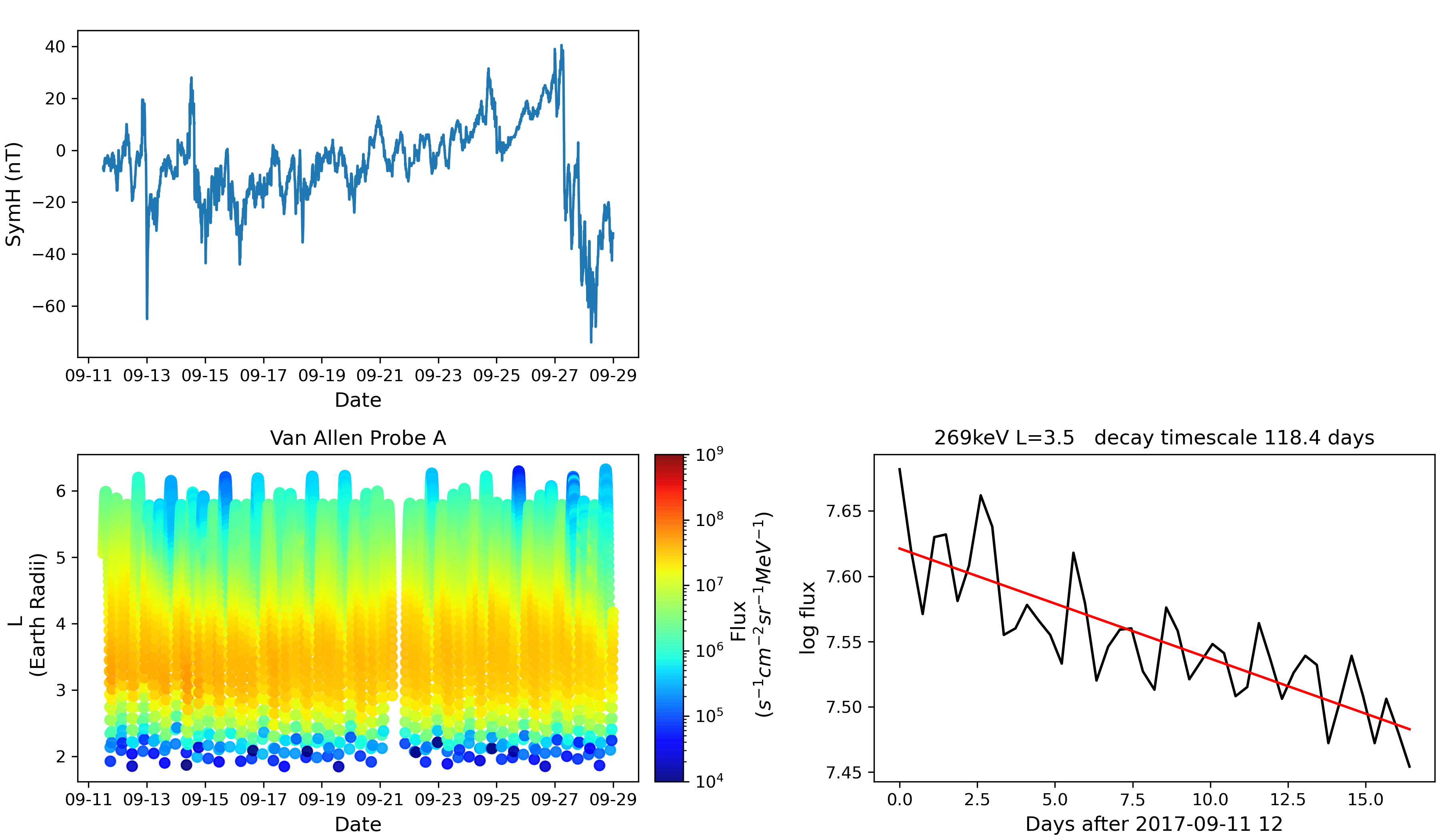}
  \caption{The same as Figure 3 but for 269 keV proton fluxes measured from 11 September to 29 September 2017. The decay timescale is only calculated at L=3.5 and is about 118 days for this case.}
  \label{fig:fig7}
\end{figure}

Figure 7 exhibits the study for 269 keV protons. The 269 keV proton fluxes generally peak around L=3.5, and are highly dynamic above L=4.0. Hence, the decay timescale is only calculated at L=3.5, which is about 118 days. 

\section{Statistical Results}
We studied 10 cases for the proton decay timescales from 27 keV to 148 keV, and produced a mean timescale and the corresponding standard deviation for each energy. The 269 keV proton fluxes are insensitive to small and moderate geomagnetic storms at around L=3.5 where the flux peaks, and we only provide one number derived from a single event shown in Figure 7. Table 1 presents a summary of the ring current proton decay timescales at their respective suited L-shells. It can be clearly seen that the proton decay timescales increase with increasing energies and increasing L-shells. The distinctly different behaviors of different energy protons can possibly be explained by the proton loss mechanism due to charge exchange. Lower energy protons have larger charge exchange cross sections, and therefore undergo more efficient charge exchanges than higher energy protons, leading to shorter lifetimes \cite{Daglis1999}. While Coulomb collision may also contribute to the loss of protons, it primarily affects low-energy protons below 10 keV \cite{Fok1991}.

The Van Allen Probe observations provide us a unique opportunity to determine the ring current decay timescales at various energies and different altitudes, and thereby provide crucial insights into the ion loss mechanisms. The statistically calculated timescales can substantially contribute to the construction of data-based ring current models. For instance, in a machine-learned ring current model which employs geomagnetic indices as the input, longer look-back windows are used for modeling higher energy protons \cite{Li2023}. 

\begin{table}
  \centering
  \begin{tabular}{llllll}
    \toprule
    Location & 27 keV & 55 keV & 99 keV & 148 keV & 269 keV \\
    \midrule
    L=5.5    &   $4.3\pm 1.2$ \\
    L=5      &   $3.8\pm 1.0$ & $4.7\pm 1.3$ & $14.7\pm2.7$ \\ 
    L=4      & & $3.1\pm 0.5$ & $8.9\pm 2.4$ & $22.2\pm4.7$ \\ 
    L=3.5    & & & &           $16.6\pm 3.2$ & $\sim118$ \\
    \bottomrule
  \end{tabular}
  \label{tab:table1}
   \caption{Statistics of proton decay timescales in the main region of the ring current. Shown in the table are mean values and standard deviations of proton decay timescale (in days) at representative energies and L shells, derived from multiple cases (except for 269 keV protons).}
\end{table}

\section{Conclusions}
By investigating the variations of ring current proton fluxes measured by Van Allen Probes, we identified energy-dependent ring current proton dynamic patterns. For instance, protons of ~10s keV energies are mainly observed at altitudes above L=4, and they are injected during each geomagnetic storm including weak ones. Protons with higher energies such as 100s keV energy are generally observed to peak at lower L shells, and they mainly respond to moderate and large storms at the center of the ring current. 

Ring current protons typically experience an exponential decay after reaching peak values during geomagnetic storms. Therefore, their decay timescale can be determined by implementing a linear regression on the logarithm of fluxes. We systematically calculated the proton decay timescales in the main region of the ring current, and the statistical results show that proton decay timescales increase with increasing energies and increasing L shells. Specifically, the decay timescale for 27 keV protons at L=5 is approximately 3.8±1 days, increasing to 4.3±1.2 days at L=5.5. The 55 keV proton decay timescale is 3.1±0.5 days at L=4, extending to 4.7±1.3 days at L=5. The 99 keV proton decay timescale is 8.9±2.4 days at L=4 and is 14.5±2.7 days at L=5. The 148 keV proton decay timescale at L=3.5 is 16.6±3.2 days, while at L=4 it is 22.2±4.7 days. The 269 keV proton decay timescale at L=3.5 is roughly 118 days. 

By studying energy-dependent proton decay timescales at varied altitudes, we gain an advanced understanding into the ring current dynamics. The proton flux decay timescales calculated from Van Allen Probe observations provide insights in developing accurate ring current models.

\section*{Acknowledgments}
JL acknowledges NASA grants LWS-80NSSC20K0201 and 80NSSC21K0522.

\bibliographystyle{unsrt}  
\bibliography{references}

\end{document}